\documentclass[conference]{IEEEtran}
\IEEEoverridecommandlockouts

\usepackage{cite}
\usepackage{amsmath,amssymb,amsfonts}
\usepackage{algorithmic}
\usepackage{graphicx}
\usepackage{textcomp}
\usepackage{xcolor}

\usepackage[utf8]{inputenc}
\usepackage{url}
\usepackage{todonotes}
\usepackage{tikz}
\usepackage{gensymb}
\usepackage{fontawesome5}
\usepackage{xspace}

\usepackage[inline, shortlabels]{enumitem}
\usepackage{tabularx}
\usepackage{booktabs}
\usepackage{pgfplots}
\usepackage{wrapfig}
\usepackage{balance}
\usepackage{caption}
\usepackage{ifthen}
\usepackage{pgfplotstable}
\usepackage{soul}
\usepackage[labelformat=simple]{subcaption}

\usepackage{tikz}
\usepackage{tkz-euclide}
\usetikzlibrary{shapes,arrows,decorations.pathreplacing,calc,arrows.meta,automata,angles,shapes,decorations,patterns,positioning,intersections,backgrounds,fit}
\usepgfplotslibrary{fillbetween}
\usetikzlibrary{trees}
\usetikzlibrary{mindmap}
\usetikzlibrary{pgfplots.groupplots}

\definecolor{grass}{HTML}{296402}
\definecolor{way}{HTML}{373737}

\definecolor{pt_blue}{RGB}{0,119,187}
\definecolor{pt_cyan}{RGB}{51,187,238}
\definecolor{pt_teal}{RGB}{0,153,136}
\definecolor{pt_orange}{RGB}{238,119,51}
\definecolor{pt_red}{RGB}{204,51,17}
\definecolor{pt_magenta}{RGB}{238,51,119}
\definecolor{pt_grey}{RGB}{187,187,187}

\usepackage{tabularx}
\usepackage{multirow}

\usepackage{cleveref} 
  \crefname{section}{Sect.}{Sects.}
  \Crefname{section}{Section}{Sections}
  \crefname{figure}{Fig.}{Figs.}
  \Crefname{figure}{Figure}{Figures} 
  \crefname{definition}{Definition}{Definitions}
  \crefname{equation}{}{}
  \Crefname{equation}{Equation}{Equations}
  \crefname{table}{Tab.}{Tabs.}
  \Crefname{table}{Table}{Tables}

\newcommand {\eg} {e.\,g.}

\newcolumntype{P}[1]{>{\centering\arraybackslash}p{#1}}
\newcolumntype{R}[1]{>{\raggedright\arraybackslash}p{#1}}

\newcolumntype{Y}{>{\centering\arraybackslash}X}

\newcommand{\error}[1][]{\ensuremath{\epsilon}\xspace}

\makeatletter
\newcommand\transformxdimension[1]{
    \pgfmathparse{((#1/\pgfplots@x@veclength)+\pgfplots@data@scale@trafo@SHIFT@x)/10^\pgfplots@data@scale@trafo@EXPONENT@x}
}
\newcommand\transformydimension[1]{
    \pgfmathparse{((#1/\pgfplots@y@veclength)+\pgfplots@data@scale@trafo@SHIFT@y)/10^\pgfplots@data@scale@trafo@EXPONENT@y}
}
\makeatother
\pgfplotsset{every tick label/.append style={font=\scriptsize}}  

\def\BibTeX{{\rm B\kern-.05em{\sc i\kern-.025em b}\kern-.08em
    T\kern-.1667em\lower.7ex\hbox{E}\kern-.125emX}}
\begin{document}

\title{A Systematic Mapping Study on Software Architecture for AI-based Mobility Systems
\thanks{This work has been funded by the Federal Ministry of Education and Research (BMBF) as part of MANNHEIM-AutoDevSafeOps (01IS22087P).}
}

\author{%
\IEEEauthorblockN{Amra Ramic}
\IEEEauthorblockA{\textit{Research Institute AImotion Bavaria} \\
\textit{Technische Hochschule Ingolstadt}\\
Ingolstadt, Germany}
\and
\IEEEauthorblockN{Stefan Kugele}
\IEEEauthorblockA{\textit{Research Institute AImotion Bavaria} \\
\textit{Technische Hochschule Ingolstadt}\\
Ingolstadt, Germany\\
Stefan.Kugele@thi.de}
}

\maketitle

\begin{abstract}
\emph{Background:}
Due to their diversity, complexity, and above all importance, safety-critical and dependable systems must be developed with special diligence. Criticality increases as these systems likely contain artificial intelligence (AI) components known for their uncertainty. As software and reference architectures form the backbone of any successful system, including \emph{safety-critical dependable systems} with \emph{learning-enabled components}, choosing the suitable architecture that guarantees safety despite uncertainties is of great eminence.
\emph{Aim:}
We aim to provide the missing overview of all existing architectures, their contribution to safety, and their level of maturity in AI-based safety-critical systems.
\emph{Method:}
To achieve this aim, we report a systematic mapping study. From a set of 1,639 primary studies, we selected 38 relevant studies dealing with safety assurance through software architecture in AI-based safety-critical systems. The selected studies were then examined using various criteria to answer the research questions and identify gaps in this area of research.
\emph{Results:}
Our findings showed which architectures have been proposed and to what extent they have been implemented. Furthermore, we identified gaps in different application areas of those systems and explained these gaps with various arguments.
\emph{Conclusion:}
As the AI trend continues to grow, the system complexity will inevitably increase, too. To ensure the lasting safety of the systems, we provide an overview of the state of the art, intending to identify best practices and research gaps and direct future research more focused.
\end{abstract}


\section{Introduction}
\label{sec:introduction}
The trend towards AI has strongly increased in recent years to the point where nearly every system now includes some form of AI component. AI has particularly made significant advances in the automotive industry. Despite AI's advantages, its application has introduced several new challenges, particularly in safety-critical intelligent mobility systems. 
As we go on to observe, a more rigorous consideration of uncertainties inherent to AI components is crucial, especially when deploying these systems in safety-critical mobility applications (\eg, \cite{yang:etal:tits23,shafaei:etal:safecomp18}). This aspect gains more relevance in the context of advanced driver-assistance systems (ADAS) and autonomous vehicles, where life-critical decisions become contingent upon AI outputs.
This paper underscores the importance of focusing on the functionality and strategic software architecture that integrates AI components into safety-critical, dependable mobility systems. These systems' functionality dictates how they cope with diverse and uncertain driving scenarios.

This study aims to broaden the current body of knowledge and provide an overview of existing solutions and approaches to how software architectures can enhance the safety of dependable AI-based systems. We conduct a \emph{systematic mapping study} to identify existing architectural solutions and assess the maturity of various application domains, thereby identifying research gaps and new research opportunities. Hence, we extend the current body of knowledge. Out of 1,693 primary studies, 38 studies were selected through a systematic voting process (cf.~\cref{fig:selection}) for detailed analysis. These studies were analysed based on various criteria (cf.~\cref{tab:exclusion}), extracting relevant data to answer the research questions posed.

Our guiding research question (RQ) is the following:
\begin{quotation}
\textbf{``How do software architectures enhance the safety of AI-driven mobility systems?''} 
\end{quotation}
With the overall aim of providing the best possible answer to this question, we have formulated the following more detailed research questions:
\begin{enumerate}[left=0pt .. 2.8\parindent,label=\textbf{RQ\arabic*}]
    \item \label{rq:1} What is the state of the art in research and practice on software architectures for AI-based mobility systems?
    \begin{enumerate}[left=0pt .. 3\parindent,label=\textbf{RQ1.\arabic*}]
    	\item \label{rq:1.1} What architectural styles were used?
    	\item \label{rq:1.2} What safety patterns were used?
    	\item \label{rq:1.3} What kind of architectural frameworks exist?
    	\item \label{rq:1.4} What architectural notations have been used?
	\end{enumerate}
	\item \label{rq:2} To what extent and how have the solutions been developed for different application areas?
    \begin{enumerate}[left=0pt .. 3.5\parindent,label=\textbf{RQ2.\arabic*}]
    	\item \label{rq:2.1} In which application areas have architectural solutions been developed?
    	\item \label{rq:2.2} What was the degree of maturity of that area?
    	\item \label{rq:2.3} What validation methods were used to evaluate the solutions?
	\end{enumerate}
		\item \label{rq:3} What patterns of AI application can be identified?
	\begin{enumerate}[left=0pt .. 3.5\parindent,label=\textbf{RQ3.\arabic*}]
    	\item \label{rq:3.1} 	What type of AI was used?
    	\item \label{rq:3.2} 	What is the relationship between AI and architectural solutions?
    	\item \label{rq:3.3} 	What is the relationship between AI and evaluation/validation methods?
    \end{enumerate}
\end{enumerate}
\paragraph{Contributions}
This work has the following scientific contributions and insights for industrial implementation.
\begin{enumerate}[(i)]
    \item An overview of existing architectural solutions, techniques, methodologies and frameworks.
    \item Identification of gaps in the research field by mapping the found literature to different research facets and application areas.
    \item A statement about the relationship between different forms of AI and the architectural solutions used.
    \item An overview of the demographic status regarding publication years, venues, and number of citations.
\end{enumerate}
\paragraph{Outline}
The remainder of this work is structured as follows: In \cref{sec:related-work}, related work is discussed, followed by the description of the technique used for the study in \cref{sec:study-design}. \Cref{sec:study-results} presents the results obtained from analysing the selected primary studies. Finally, \cref{sec:discussion} discusses the results obtained while \cref{sec:conclusion} concludes this paper.
\section{Related Work}
\label{sec:related-work}
Safety-critical systems are no longer a new area of research. Even a simple search on this term will yield many different reviews and studies reflecting the current state of research. Existing studies cover general topics, such as those discovering the challenges and directions in the development of safety-critical systems \cite{knight:icse02}, as well as more specific topics, such as exploring the ethics of these systems \cite{bowen:acm00}. Almost all such systems have software components whose safety must be ensured.
Due to the number and variety of systems, different architectures emerged. Guessi~et~al.~\cite{guessi:etal:12} focused only on the description of architectures and conducted a study on how to represent software architectures and reference architectures for embedded systems. As a result, they gained a detailed overview and found that different approaches can be proposed and used, with no consensus on better representing embedded system architectures. The same was done three years later also by Guassi~et~al.~\cite{guessi:etal:15} for Systems of Systems (SoS). Because of the similar outcome, the authors have provided hints on how a consensus might be reached and what aspects of SoS architectural descriptions need further investigation. Klein~\cite{klein:qosa:13} has extended the field of research to examine not only the descriptions but also the architectures of systems of systems in general. 
Regardless of the architecture chosen, according to Arshad and Usman~\cite{arshad:ease:11}, it should be noted that non-functional requirements (NFRs) are strongly influenced by the architectural decisions and designs made in the architecture phase. 
They studied the quantity and type of work reported on security as an NFR at the software architecture level. The results show that most of the work in this area is about proposing and evaluating a solution, indicating research gaps. A more general study has been conducted by Martínez-Fernández~et~al.~\cite{martinez:22}, which examines software engineering approaches to building, operating, and maintaining AI-based systems. Our approach builds on the same idea of studying safety at the architectural level as NFR, but only for systems with embedded AI components.
\section{Study Design}
\label{sec:study-design}
This section describes the techniques used to search, classify, and select relevant studies for further analysis. We followed Kitchenham's~\cite{kitchenham:07} guidelines for conducting systematic literature reviews and mapping studies. 
\subsection{Search Strategy}
To find relevant publications, we began by selecting certain keywords related to our topic. The chosen keywords are: ``Software Architecture,'' ``Software Design,'' ``Safety,'' ``Artificial Intelligence,'' ``Deep Learning,'' and ``Machine Learning.''. We used both ``Design'' and ``Architecture'' for the architecture keyword due to their frequent equivalence and overlap. While AI, ML, and DL are distinct terms, they are often used interchangeably. Therefore, we included all three, even though AI is the broadest concept.
\begin{table}
    \centering
    \caption{Search Strings}
    \label{tab:search_strings}
    \begin{tabularx}{\columnwidth}{@{}X@{}}
        \toprule
        \textbf{Digital Library} and \textbf{Search String}\\
        \midrule
        \emph{ACM Digital Library} \\
        "Software Architecture" AND "Safety" AND ("Artificial Intelligence" OR "AI" OR "Machine Learning" OR "ML" OR "Deep Learning" OR "DL")\\
        \midrule
        \emph{IEEE Xplore} \\
        "Full Text Only":"Software Architecture" AND "Full Text Only":"Safety" AND ("Full Text Only":"AI" OR "Full Text Only":"Artificial Intelligence" OR "Full Text Only":"Machine Learning" OR "Full Text Only":"ML" OR "Full Text Only":"Deep Learning" OR "Full Text Only":"DL")\\
        Filter: mobile robots, software architecture, learning (artificial intelligence), traffic engineering computing, embedded systems, software engineering, safety-critical software, road vehicles, road safety, autonomous aerial vehicles, automobiles, real-time systems, collision avoidance, decision making\\
        \midrule
        \emph{DBLP} \\
        (Software Architecture $\vee$ Software Design $\vee$ Architecture $\vee$ Design) $\wedge$ Safety $\wedge$ (Artificial Intelligence $\vee$ AI $\vee$ Deep Learning $\vee$ DL $\vee$ Machine Learning $\vee$ ML)\\
        \bottomrule
    \end{tabularx}
\end{table}
After identifying the relevant keywords, the next step was to create search strings that capture the essence of our research. We applied the combined search strings to the three most commonly used digital libraries in software engineering, namely ACM~DL, IEEE~Xplore, and DBLP. To streamline the search process, we have listed the search strings for each digital library in \cref{tab:search_strings}.
\subsection{Relevant Papers Selection}
After applying the search strings, we found 1,639 primary studies containing the keywords in the title, abstract, or full text. However, we found duplicates in different databases and thus removed 119 duplicate studies. This left us with 1,520 distinct studies for detailed examination. We aimed to reduce this number further using the exclusion criteria (\cref{tab:exclusion}). We excluded studies that were not written in English or inaccessible for download. Additionally, we eliminated most studies due to a lack of relevance in content or application domains, such as medicine. The studies we selected for our study had to meet the inclusion criteria listed in \cref{tab:exclusion}. On the one hand, the primary study had to contain the specified keywords either in the title, abstract, or body of the study. On the other hand, it must have been published in the mobility domain such as \eg, aerospace or automotive.
\begin{table}
    \centering
    \caption{Inclusion and Exclusion Criteria}
    \label{tab:exclusion}
    \begin{tabularx}{\columnwidth}{@{}R{1em}X@{}}
    \toprule
    \textbf{No.} & \textbf{Inclusion Criteria}\\
    \midrule
    1. & Title, keywords, abstract, or the full-text are related to software architecture of AI-based mobility systems\\
    2. & The paper is related to automotive, aerospace, or robotics\\
    \toprule
    \textbf{No.} & \textbf{Exclusion Criteria}\\
    \midrule
    1. & Paper not related to software architecture for AI-based mobility systems\\
    2. & Paper is not written in English\\
    3. & Paper is not available to download\\
    4. & Front matter, back matter, proceedings descriptions, full proceedings\\
    5. & Preprints, if there is a published version in a journal or conference\\
    6. & Paper is a duplicate\\
    \bottomrule
\end{tabularx}    
\end{table}
The exclusion process in this study was meticulously conducted over three iterative voting rounds by the paper's authors, comprising a senior researcher and a research associate. To ensure objectivity, Cohen's kappa \cite{cohen:60} was employed throughout to quantify the agreement level between the voters.

In the first round (each author either voted for or against the paper), we evaluated 1,520 study titles, identifying 230 relevant primary studies. This round achieved a Cohen's kappa agreement score of 0.70, indicating substantial concordance per the predefined scale. After the title review, the second round scrutinised the abstracts of studies whose eligibility could not be determined based solely on their titles. This led to further exclusions, narrowing the selection to 166 studies. The agreement score from this round was an impressive 0.86, indicating almost perfect alignment between the voters.

\begin{figure}
    \centering
    \resizebox{.9\linewidth}{!}{
    \begin{tikzpicture}[%
    node distance=1.5em and 0.5em,
    node/.style={draw, minimum height=2.3em, minimum width=3em, text width=3em, align=center, rounded corners = 2pt, font=\footnotesize\sffamily}     ,
    database/.style={cylinder, shape border rotate=90, draw, text width=5em, minimum height=2em, shape aspect=.25, align=center, font=\footnotesize\sffamily}
    ]
    \node[node, text width=22em, text centered, fill=gray!30!white](ss) {Search strings};
%
    \node [database, below =of ss.south west, anchor=north west] (acmdl) {ACM DL\\(723)};
    \node (dblp) [database, below=of ss.south east, anchor=north east] {DBLP\\(49)};
    \node (ieeexplore) [database, at=($(acmdl)!0.5!(dblp)$), text width=5em] {IEEE Xplore (867)};
    \draw [<-](acmdl) -- (ss.south -| acmdl.north);
    \draw [->](ss) -- (ieeexplore);
    \draw [<-](dblp) -- (ss.south -| dblp.north);
%
    \node[node, below left=of acmdl, fill=gray!10!white, text width=6em](dupes) {Duplicate removal};
    \node[node, minimum width=1cm, below=of ieeexplore](dupesNr){1,520};
    \draw [->](acmdl) |- (dupesNr);
    \draw [->](ieeexplore) -- (dupesNr);
    \draw [->](dblp) |- (dupesNr);
%
    \node[node, minimum width=1cm, below=of dupesNr](title){230};
    \node[node, below=of dupes, fill=gray!10!white, text width=6em](vtitle) {Voting based\\on title};
    \draw [->](dupesNr) -- (title);
 %
    \node[node, minimum width=1cm, below=of title](abstract){166};
    \node[node, below=of vtitle, fill=gray!10!white, text width=6em](vabstract) {Voting based\\on abstract};
    \draw [->](title) -- (abstract);
%
    \node[node, minimum width=1cm, below =of abstract](full){141};
    \node[node, minimum width=1cm, below =of vabstract, text width=6em, fill=gray!10!white](vfull) {Voting based on full text};
    \draw [->](abstract) -- (full);
    \draw [->](vabstract) -- (vfull);
%
    \node[node, minimum width=1cm, below =of full](exclusion){38};
    \node[node, minimum width=1cm, below =of vfull, text width=6em, fill=gray!10!white](vexclusion) {Relevance for RQs};
    \draw [->](full) -- (exclusion);
    \draw [->](vfull) -- (vexclusion);
    \node[node, minimum width=1cm, right =3em of exclusion, text width=6em, fill=gray!30!white](total) {Total};
    \draw [->](exclusion) -- (total);
    \draw [->](dupes) -- (vtitle);
    \draw [->](vtitle) -- (vabstract);
    \end{tikzpicture}}%
    \caption{Summary of the selection procedure}
    \label{fig:selection}
\end{figure}
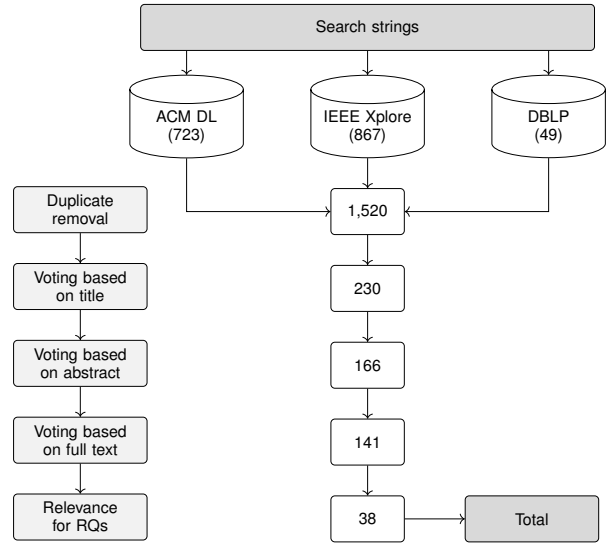

For the remaining studies, a comprehensive review of the full text was undertaken in the final voting round, which continued until a perfect agreement score of 1.00 was achieved, culminating in a final selection of 141 primary studies. \Cref{fig:selection} provides a detailed graphical representation of the voting outcomes across the rounds.
Notably, despite some studies initially advancing to the final round based on title or abstract assessments, additional scrutiny during the full-text reviews led to their exclusion if they were found not to address the research questions adequately. Ultimately, detailed responses to the research questions were extracted from 38 studies.
\subsection{Data Extraction and Classification Scheme}
Each selected study was closely analysed, and relevant data was extracted to answer the research questions. Please refer to \cref{tab:extraction} to see what data we needed to extract and for which research questions.
\begin{table}[t!]
    \centering
    \caption{Data Extraction}
    \label{tab:extraction}
    \begin{tabularx}{\columnwidth}{@{}R{2em}R{7.4em}X@{}}
    \toprule
    \textbf{RQ} & \textbf{Information} & \textbf{Description}\\
    \midrule
    RQ1.1 & Architecture type & Proposed architecture or a design pattern \\
    RQ1.2 & Architectural framework & Architectural framework used \\
    RQ1.3 & Architecture notation & Notation language used for the graphical or textual representation \\
    RQ2.1 & Research facet & The classification of a paper according to \cite{wieringa:06}\\
    RQ2.2 & Validation method & The method the proposed approach was validated or evaluated \\
    RQ2.3 & Application area & The application domain \\
    RQ3.1 & Type of AI & The form in which the AI occurs \\
    \bottomrule
    \end{tabularx}
\end{table}

\section{Study Results}
\label{sec:study-results}
\begin{figure}
    \centering
    \pgfplotstableread[row sep=\\,col sep=&]{
    pattern & number\\
    Redundancy  &   10\\
    Monitoring  &   4\\ 
    Hypervisor  &   3\\
    Watchdog    &   2\\
    TMR &   1\\
    Fault recovery  &   1\\
    Fault detection &   1\\
    N-self checking &   1\\
    Sanity checking &   1\\
    Predict. DNN inference &   1\\
}\mydata
\begin{tikzpicture}
    \begin{axis}[
            ybar,
            symbolic x coords={Redundancy, Monitoring, Hypervisor,Watchdog, TMR,  Fault recovery, Fault detection, N-self checking,Sanity checking, Predict. DNN inference},
            xtick=data,
            ytick={0,2,4,6,8,10},
            ymin=-0.5,
            grid, grid style=dashed,
            xticklabel style={rotate=45,anchor=east,align=center},
            nodes near coords,
            nodes near coords align={vertical},
            nodes near coords style={font=\small, text=black, fill=white}, 
            width=\columnwidth,
            height=5cm,
            enlarge y limits=0.2
        ]
        \addplot [fill=pt_blue]table[x=pattern,y=number]{\mydata};
    \end{axis}
    \end{tikzpicture}
    \caption{Safety Patterns}
    \label{fig:pattern}
\end{figure}
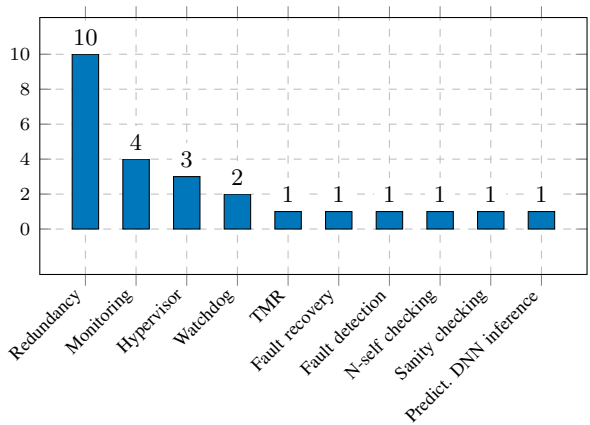

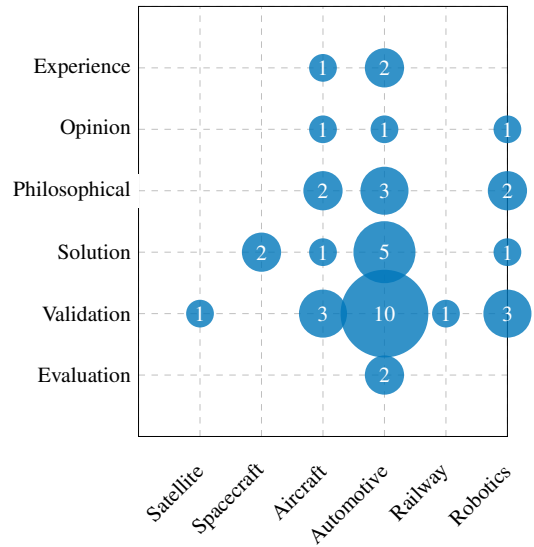
\begin{figure}
    \centering
    \resizebox{.8\linewidth}{!}{
    \begin{tikzpicture}
    \tikzset{help lines/.style={dashed, gray!50}}
    \def\pHeight{0.5\columnwidth}
	\def\pWidth{\columnwidth}
	\def\cat{6}
	\def\foo{7}
    \foreach \c in {0,1,...,\cat} {
        \draw[help lines] (\c,0) -- (\c,\foo);
    }
    \foreach \c in {0,1,...,\foo} {
        \draw[help lines] (0, \c) -- (\cat,\c);
    }
    \foreach \l [count=\c from 1] in {{Satellite}, {Spacecraft}, {Aircraft}, {Automotive}, {Railway}, {Robotics}} {
        \node [rotate=45,anchor=east,align=right, inner sep=0pt] at (\c, -0.5) {\l};
    }
    \draw[black] (0, 0) rectangle (\cat,\foo);
    \tikzstyle{stuff_fill}=[rectangle,fill=white!20]
    \foreach \l [count=\c from 1] in {{Evaluation}, {Validation}, {Solution}, {Philosophical}, {Opinion}, {Experience}} {
        \node[stuff_fill, anchor=east] at (0, \c) {\l};
    }
    \pgfplotstableread{data/appl-area.dat}\table
    \pgfplotstablegetrowsof{\table}
    \pgfmathsetmacro{\M}{\pgfplotsretval-1}
    \pgfplotstablegetcolsof{\table}
    \pgfmathsetmacro{\N}{\pgfplotsretval-1}
    \foreach \row in {0,...,\M}{
        \foreach \col in {0,...,\N}{
            \pgfplotstablegetelem{\row}{[index]\col}\of\table
                \ifnum\col=0
                    \xdef\x{\pgfplotsretval}
                \fi
                \ifnum\col=1
                    \xdef\y{\pgfplotsretval}
                \fi
                \ifnum\col=2
                    \xdef\radius{\pgfplotsretval}
                \fi
        }
        \definecolor{mycolor}{RGB}{\pdfuniformdeviate 256,%
            \pdfuniformdeviate 256,%
            \pdfuniformdeviate 256}
        \fill[pt_blue,opacity=.8] (\x,\y)circle({sqrt(\radius/3.1415)*0.4});
        \node[text = white] at (\x,\y) {\radius};
    }
\end{tikzpicture}}
\caption{Application areas and their maturity}
\label{fig:appl-area}
\end{figure}

The selection process has resulted in 38 papers.\footnote{In case of acceptance, the papers will be mentioned and references added.} This section details the data analysis findings and how they answer the research questions. Please note that the numbers in the charts below may not add up to exactly 38 since a primary study may address multiple aspects or no aspect at all.
%
\paragraph*{\ref{rq:1.1}}
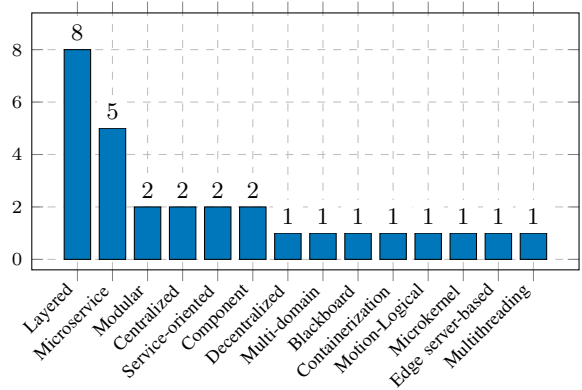
\begin{figure}
    \centering
    \pgfplotstableread[row sep=\\,col sep=&]{
    architecture & number \\
Layered    &   8\\
Microservice    &   5\\
Modular    &   2\\
Centralized    &   2\\
Service-oriented    &   2\\
Component    &   2\\
Decentralized    &   1\\
Multi-domain    &   1\\
Blackboard    &   1\\
Containerization   &   1\\
Motion-Logical    &   1\\
Microkernel    &   1\\
Edge server-based    &   1\\
Multithreading    &   1\\
}\mydata
\begin{tikzpicture}
    \begin{axis}[
            ybar,
            symbolic x coords={Layered,
            Microservice,
            Modular,
            Centralized,
            Service-oriented,
            Component,
            Decentralized,
            Multi-domain,
            Blackboard,
            Containerization,
            Motion-Logical,
            Microkernel,
            Edge server-based,
            Multithreading,
},
            xtick=data,
            grid, grid style=dashed,
            xticklabel style={rotate=45,anchor=east,align=center},
            nodes near coords,
            nodes near coords align={vertical},
            nodes near coords style={font=\small, text=black, fill=white}, 
            width=\columnwidth,
            height=5cm,
            enlarge y limits=0.2
        ]
        \addplot [fill=pt_blue]table[x=architecture,y=number]{\mydata};
    \end{axis}
    \end{tikzpicture}
    \caption{Software architectures}
    \label{fig:architecture}
\end{figure}
While analysing the papers, we observed various architectural styles presented in \cref{fig:architecture}. Most of the solutions used a layered architecture, like the three-layer or multi-layer architecture, which is unsurprising because it is one of the most commonly used styles. The micro-services architecture was also widespread, especially since AI components are often integrated as separate and independent functions. Service-oriented and micro-service architectures are best suited for such cases. We identified other architectures that were represented relatively evenly, including well-known architectures such as service-oriented or component architectures and rare ones such as edge server-based or motion-logical architectures.
%
\paragraph*{\ref{rq:1.2}}
We found that safety patterns are reusable templates that can be used to solve common problems related to the safety of AI-based systems. During our analysis, we discovered several patterns commonly used to ensure the safety of AI-based mobility systems. These patterns are illustrated in \cref{fig:pattern}. One of the most popular methods is redundancy, where multiple components are designed to perform the same function. There are several types of redundancy, including heterogeneous, homogeneous, or on-demand redundancy. Due to the uncertainty of AI, redundancy is frequently used to ensure the accuracy of the output generated by the AI. This is achieved by using one or more AI components that are either identically or differently implemented. Additionally, the pattern of runtime or safety monitoring is also commonly used. This involves continuous monitoring of the AI components' functionality to trigger an appropriate safety mechanism in case of a malfunction. Although not as commonly used, other patterns such as watchdog, hypervisor, or self-checking are still viable solutions.
%
\paragraph*{\ref{rq:1.3}}
Our investigation showed that three of the evaluated frameworks---AUTOSAR, Apollo AD, and OpenDLV---are specifically designed for the automotive industry, focusing on ensuring functional safety. This is especially important in the field of perception for autonomous driving, where models like Deep Neural Networks (DNNs) and Convolutional Neural Networks (CNNs) are commonly used. It is essential to ensure the reliability of these models to ensure vehicular safety. This observation is in line with the findings of our study (\ref{rq:2.1}), which are depicted in \cref{fig:appl-area}. Additionally, our study also examined self-adaptive systems, with a particular emphasis on the MAPE (Monitor-Analyse-Plan-Execute) framework or its extended version, MAPE-K (K = over a shared Knowledge), which incorporates a learning component.
\begin{figure}
    \centering
        \pgfplotstableread[row sep=\\,col sep=&]{
    framework & number \\
    AUTOSAR & 1\\
    OpenDLV & 1\\
    SPHERE & 1\\
    compage-icom & 1\\
    Apollo AD & 1\\
    MAPE-K & 2\\
    MAPE & 1\\
    ADTF & 1\\
    ANUNNAKI & 1\\
    None & 28\\
}\mydata
\begin{tikzpicture}
    \begin{axis}[
            ybar,
            symbolic x coords={AUTOSAR,OpenDLV,SPHERE,compage-icom,Apollo AD,MAPE-K,MAPE,ADTF,ANUNNAKI,None
},
            ytick={0,5,10,15,20,25,30},
            xtick=data,
            grid, grid style=dashed,
            xticklabel style={rotate=45,anchor=east,align=center},
            nodes near coords,
            nodes near coords align={vertical},
            nodes near coords style={font=\small, text=black, fill=white}, 
            width=\columnwidth,
            height=5cm,
            enlarge y limits=0.2
        ]
        \addplot [fill=pt_blue]table[x=framework,y=number]{\mydata};
    \end{axis}
    \end{tikzpicture}
    \caption{Architectural Frameworks}
    \label{fig:frameworks}
\end{figure}
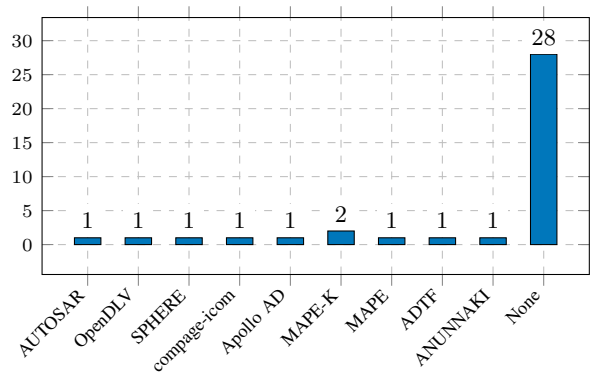
%
\paragraph*{\ref{rq:1.4}}
In our analysis of architectural notations used in the proposed solutions, we categorised them into three distinct groups. The first group comprises informal notations, which depict architecture through syntax-free, primarily graphical elements such as boxes and arrows. The second group includes semi-formal notations, which facilitate the description of the architecture, design, and implementation of complex software systems. One notable example of semi-formal notations is the Unified Modelling Language (UML), which is standardised. The third group encompasses formal notations that are tailored specifically for software architectures. One such example is the Architecture Description Language (AADL), which is crucial for modelling and analysing real-time, safety-critical, and embedded systems \cite{hugues} and Markov Decision Process (MDP), which follows established mathematical methodologies.

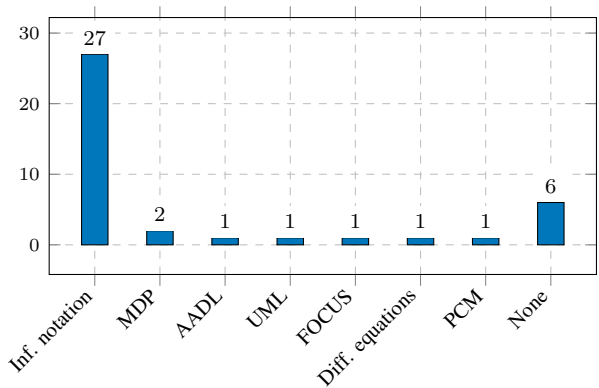
\begin{figure}
    \centering
    \begin{tikzpicture}
        \begin{axis}[
            ybar,
            symbolic x coords={Inf. notation, MDP, AADL, UML, FOCUS, Diff. equations, PCM, None},
            xtick={Inf. notation, MDP, AADL, UML, FOCUS, Diff. equations, PCM, None},
            xticklabels={Inf. notation, MDP, AADL, UML, FOCUS, Diff. equations, PCM, None},
            grid, grid style=dashed,
            nodes near coords,
            nodes near coords align={vertical},
            nodes near coords style={font=\footnotesize, text=black, fill=white}, 
            ybar=-0.35cm,
            xticklabel style={rotate=45,anchor=east,align=center,font=\footnotesize,},
            width=\columnwidth,
            height=5cm,
            enlarge y limits=0.2
        ]
            \addplot[fill=pt_blue]table[x=notation, y=number]{\mydata};
        \end{axis}
    \end{tikzpicture}
    \caption{Used architectural notations}
    \label{fig:notation}
\end{figure}

Furthermore, some of the studies examined did not employ any architectural representation. According to the data visualised in \cref{fig:notation}, just over 10\% of the studies utilised formal or semi-formal notations. More than half relied on informal notations, and the remainder did not use any notation. While the predominance of informal notations might suffice within a community of software architects within a specific domain, mobility systems' growing complexity and critical nature necessitate interdisciplinary collaboration. In such contexts, employing formal or at least semi-formal notations is advantageous. These notations bridge communication gaps among various stakeholders and across different domains, thus enhancing collaborative efforts and understanding in developing complex systems.
\pgfplotstableread[row sep=\\,col sep=&]{
    notation & number\\
    Inf. notation    &   27\\
    MDP & 2\\
    AADL    &	1\\
    UML	&   1\\
    FOCUS   &	1\\
    Diff. equations & 	1\\
    PCM &   1\\
    None    &	6\\
}\mydata
%
\paragraph*{\ref{rq:2.1} and \ref{rq:2.2}}
In this section, we have integrated two primary research questions: firstly, the application areas where the studies were conducted, and secondly, the maturity level of these areas. Our focus is primarily on mobility systems, particularly within the automotive industry. Given its extensive scope, we specifically targeted the automotive sector to refine our research scope. This sector itself is quite diverse, encompassing various vehicle types, including air and ground underground vehicles. Based on this categorisation, we constructed the application areas depicted in \cref{fig:appl-area}. The analysis reveals a significant concentration of research within the automotive, particularly the car industry, with a substantial volume of studies also targeting the aircraft sector. Conversely, other sectors are represented to a much lesser extent.

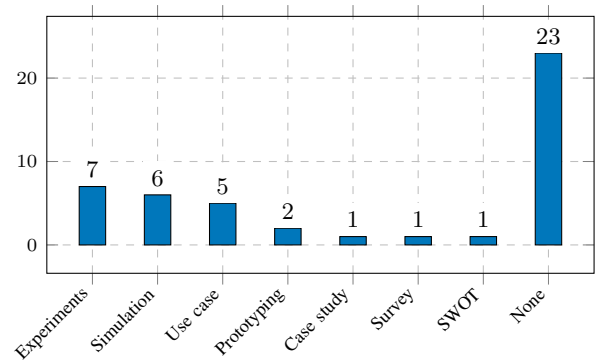
\begin{figure}
    \centering
    
\pgfplotstableread[row sep=\\,col sep=&]{
    method & number \\
    Experiments	&   7\\
    Simulation	&   6\\
    Use case	&   5\\
    Prototyping	&   2\\
    Case study	&   1\\
    Survey	&   1\\
    SWOT	&   1\\
    None	&   23\\
}\mydata

\begin{tikzpicture}
    \begin{axis}[
            ybar,
            symbolic x coords={Experiments,Simulation,Use case,Prototyping,Case study,Survey,SWOT,None},
            xtick=data,
            grid, grid style=dashed,
            xticklabel style={rotate=45,anchor=east,align=center},
            nodes near coords,
            nodes near coords align={vertical},
            nodes near coords style={font=\small, text=black, fill=white}, 
            width=\columnwidth,
            height=5cm,
            enlarge y limits=0.2
        ]
        \addplot [fill=pt_blue]table[x=method,y=number]{\mydata};
    \end{axis}
    \end{tikzpicture}
\caption{Validation/Evaluation methods}
\label{fig:validation}
\end{figure}

To assess a field's maturity, we adopted the classification framework proposed by Wieringa~et~al.~\cite{wieringa:06}, which delineates six distinct facets of scholarly works and allows for a uniform classification. An area is mature if primary studies address diverse research type facets.
\begin{itemize}
    \item \emph{Evaluation paper}: Techniques are implemented and evaluated in a large-scale industrial, academic, or other real-world setting.
    \item \emph{Validation paper}: Investigated techniques are novel and have not yet been implemented on a large scale in an industrial or academic setting.
    \item \emph{Solution proposal}: A solution to an issue is suggested, whether novel or an expansion of an existing approach.
    \item \emph{Philosophical paper}: It proposes a new way of looking at existing problems by restructuring the field into a taxonomy, conceptual framework, or literature review.
    \item \emph{Opinion paper}: The authors present their opinion on a problem area with a critical view.
    \item \emph{Experience paper}: The authors provide a retrospective analysis of their experience in developing, applying, and evaluating a technique to improve engineering processes.
\end{itemize}
In this analysis, there is a pronounced emphasis on solution and validation papers, suggesting that while many publications propose innovative approaches, their validation often occurs at a modest scale, such as through simulations or prototypes. This suggests a relative immaturity in the field and underscores the need for further validated approaches within industry settings. The categories represented in the upper half of \cref{fig:appl-area} are designed to offer either a subjective (opinion-based) or objective (philosophical) perspective on one or more existing solutions. The sparse density in this upper segment highlights a \emph{research gap}. This gap is particularly significant in spacecraft and railway systems, where papers are critically needed to analyse and synthesise existing solutions. We are confident this study will contribute to bridging this gap.
%
\paragraph*{\ref{rq:2.3}}
Irrespective of the chosen architectural solution or pattern, its correctness and functionality remain unproven. Any proposed approach should be validated, at a minimum, through a small use case or experiment. As depicted in \cref{fig:validation}, over half of the approaches lack validation or evaluation, presenting significant risks to users who implement these unverified methods. Furthermore, the most prevalent methods, such as simulation and experimentation, do not constitute validation of the actual system. Therefore, a notable gap exists in evaluating the approach and its outcomes on real systems.
%
\paragraph*{\ref{rq:3.1}}
Analysis of the primary studies revealed the distribution of various AI technologies. Please note that the primary studies were often not very precise in classifying the technologies used. We have, therefore, tried to be as precise as possible. However, we know this categorisation leads to hierarchical inclusion relationships (\eg, ML is an AI method, and DL is an ML method). Many papers focus on artificial intelligence (15) and machine learning (13), which initially seems justifiable. However, it prompts the question of why there are significantly fewer studies on reinforcement learning (2), CNNs (3), or DNNs (8). As illustrated in \cref{tab:citations}, the most cited studies predominantly originate from the automotive sector, where autonomous driving frequently emerges in discussions about integrating AI with automotive technology.
Given that CNNs are adept at image recognition, it is puzzling why they are not more frequently applied to traffic sign or pedestrian recognition. Similarly, reinforcement learning, which could optimally serve self-driving cars due to its ability to learn from mistakes via a punishment and reward system and adapt to dynamic environments, is also underutilised. This discrepancy suggests a potential area for further research to explore why these technologies are underrepresented and what unique solutions they could offer.
%
\paragraph*{\ref{rq:3.2}}
To further address research question \ref{rq:1.1}, we examined how individual architectures are applied to different AI technologies. Our analysis shows that ML solutions are predominantly integrated using microservices architectures. This approach allows the ML applications to function independently from the main software product, offering two main benefits:
(i) \emph{Distributed Development:} Facilitates faster market entry and significantly enhances software scalability.
(ii) \emph{Decentralised Responsibilities:} While distributing tasks across services complicates the data collection process for AI learning, this issue can be mitigated by adopting a monolithic architecture. This allows for centralised management of all services, simplifying data transfer across different layers. Consequently, layered architecture is prevalent across nearly all AI implementations.

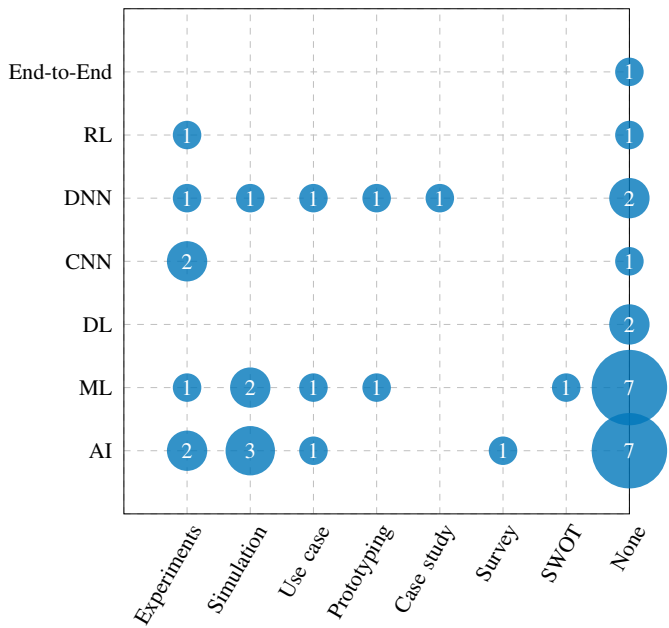
\begin{figure}
    \centering
    \resizebox{\linewidth}{!}{
\begin{tikzpicture}
    \tikzset{help lines/.style={dashed, gray!50}}
    \def\pHeight{0.5\columnwidth}
	\def\pWidth{\columnwidth}
	\def\cat{8}
	\def\foo{8}

    \foreach \c in {0,1,...,\cat} {
        \draw[thin,help lines] (\c,0) -- (\c,\foo);
    }
    \foreach \c in {0,1,...,\foo} {
        \draw[thin,help lines] (0, \c) -- (\cat,\c);
    }
    
    \draw[black] (0, 0) rectangle (\cat,\foo);
    \foreach \l [count=\c from 1] in {{Experiments},{Simulation},{Use case},{Prototyping},{Case study},{Survey},{SWOT}, {None}} {
        \node [rotate=60, anchor=east, inner sep=0pt, xshift=10pt] at (\c, -0.5) {\l};
    }

    \tikzstyle{stuff_fill}=[rectangle,fill=white!20]
  
    \foreach \l [count=\c from 1] in {{AI}, {ML}, {DL}, {CNN}, {DNN}, {RL}, {End-to-End}, } {
        \node[stuff_fill, anchor=east, inner sep=0pt, xshift=-5pt] at (0, \c) {\l};
    }
  
    \pgfplotstableread{data/validation-ai.dat}\table
    \pgfplotstablegetrowsof{\table}
    \pgfmathsetmacro{\M}{\pgfplotsretval-1}
    \pgfplotstablegetcolsof{\table}
    \pgfmathsetmacro{\N}{\pgfplotsretval-1}
  
    \foreach \row in {0,...,\M}{
        \foreach \col in {0,...,\N}{
            \pgfplotstablegetelem{\row}{[index]\col}\of\table
                \ifnum\col=0
                    \xdef\x{\pgfplotsretval}
                \fi
                \ifnum\col=1
                    \xdef\y{\pgfplotsretval}
                \fi
                \ifnum\col=2
                    \xdef\radius{\pgfplotsretval}
                \fi
        }
        \definecolor{mycolor}{RGB}{\pdfuniformdeviate 256,%
            \pdfuniformdeviate 256,%
            \pdfuniformdeviate 256}
        \fill[pt_blue,opacity=0.8] (\x,\y)circle({sqrt(\radius/3.1415)*0.4});
        \node[color=white] at (\x,\y) {\radius};
    }
\end{tikzpicture}}
\caption{The correlation between architectures and AI types.}
\label{fig:validation-ai}
\end{figure}

Interestingly, end-to-end learning-based mobility systems typically employ a component architecture. This is particularly evident in autonomous driving applications, where data from one component, like Lidar, is processed by other components for feature extraction, such as object detection. After data processing and feature extraction, a model of the environment is created within a central component. The system then evaluates this model to determine the necessary actions, which are executed based on system policy and vehicle objectives, selecting the optimal behaviour through optimisation criteria.
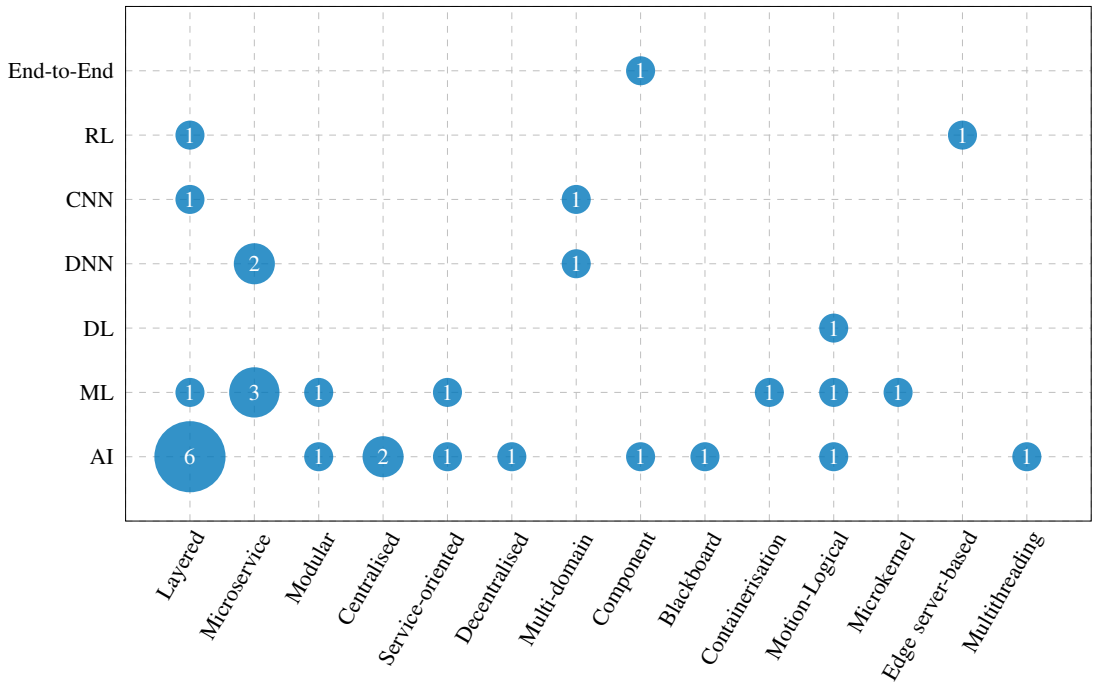
\begin{figure*}
    \centering
    \resizebox{0.8\linewidth}{!}{
\begin{tikzpicture}
    \tikzset{help lines/.style={dashed, gray!50}}
    \def\pHeight{0.1\textwidth}
	\def\pWidth{.4\textwidth}
	\def\cat{15}
	\def\foo{8}

    \foreach \c in {0,1,...,\cat} {
        \draw[thin,help lines] (\c,0) -- (\c,\foo);
    }
    \foreach \c in {0,1,...,\foo} {
        \draw[thin,help lines] (0, \c) -- (\cat,\c);
    }
    
    \draw[black] (0, 0) rectangle (\cat,\foo);
    
    \foreach \l [count=\c from 1] in {{Layered},{Microservice},{Modular},{Centralised},{Service-oriented},{Decentralised}, {Multi-domain}, {Component},{Blackboard},{Containerisation},{Motion-Logical},{Microkernel},{Edge server-based},{Multithreading}
} {
        \node [rotate=60, anchor=east, inner sep=0pt, xshift=10pt] at (\c, -0.5) {\l};
    }

    \tikzstyle{stuff_fill}=[rectangle,fill=white!20]
  
    \foreach \l [count=\c from 1] in {{AI}, {ML}, {DL}, {DNN}, {CNN},{RL}, {End-to-End}} {
        \node[stuff_fill, anchor=east, inner sep=0pt, xshift=-5pt] at (0, \c) {\l};
    }
  
    \pgfplotstableread{data/architecture-ai.dat}\table
    \pgfplotstablegetrowsof{\table}
    \pgfmathsetmacro{\M}{\pgfplotsretval-1}
    \pgfplotstablegetcolsof{\table}
    \pgfmathsetmacro{\N}{\pgfplotsretval-1}
  
    \foreach \row in {0,...,\M}{
        \foreach \col in {0,...,\N}{
            \pgfplotstablegetelem{\row}{[index]\col}\of\table
                \ifnum\col=0
                    \xdef\x{\pgfplotsretval}
                \fi
                \ifnum\col=1
                    \xdef\y{\pgfplotsretval}
                \fi
                \ifnum\col=2
                    \xdef\radius{\pgfplotsretval}
                \fi
        }
        \definecolor{mycolor}{RGB}{\pdfuniformdeviate 256,%
            \pdfuniformdeviate 256,%
            \pdfuniformdeviate 256}
        \fill[pt_blue,opacity=0.8] (\x,\y)circle({sqrt(\radius/3.1415)*0.4});
        \node[color=white] at (\x,\y) {\radius};
    }
\end{tikzpicture}}
\caption{The correlation between software architectures and the contained AI types.}
\label{fig:architecture-ai}
\end{figure*}
%
\paragraph*{\ref{rq:3.3}}
Even though more than half of the architectural solutions have not been validated or evaluated, \cref{fig:validation-ai} highlights a significant gap. Notably, there is a dearth of verification concerning the effectiveness of deep learning and end-to-end learning approaches, given their limited representation in the literature. Additionally, AI applications, particularly those involving machine learning, generally undergo validation through experiments or simulations. It is important to note that only one application involving DNNs has been assessed within an industrial case study. Previous discussions in question \ref{rq:2.3} have addressed the methods of validation and evaluation used; however, it is evident that there is a widespread lack of approaches evaluated within an industrial context, regardless of the AI technology employed.
%
\paragraph*{Extended Data Collection}
In addition to addressing the research questions, we collected supplementary information about the primary studies. Our analysis aimed to chart AI-based mobility system research trends over time and across different venues. The initial study emerged from a conference in 2005, but the ensuing 15 years saw minimal activity, averaging only one publication every five years. A significant uptick occurred in 2019, with nearly 90\% of the published studies, particularly in 2020. For further details, please refer to \cref{fig:trend}.
We also sought to understand the reasons behind this trend. AI technology has been present since the 1990s, with a notable early implementation in 2004 through DARPA’s Grand Challenge~\cite{ozguner:07}, which spurred American innovation in autonomous vehicle technologies for military applications. This challenge fostered a research community that, a decade later, has been pivotal in advancing autonomous vehicles into viable, near-term realities. Initially, the focus was on proving the feasibility of autonomous driving with AI. More recently, as autonomous driving has become more tangible, the emphasis has shifted towards enhancing safety and social acceptability of mobility systems in the presence of AI. The study's venues and trends over the years are depicted in \cref{fig:trend} with the most cited primary studies listed in \cref{tab:citations}.

We anticipate that the prevalence and impact of AI in mobility systems will continue to expand significantly.
\begin{figure}
    \centering
\begin{tikzpicture}
\begin{groupplot}[
    group style={
        group name=trend-years,
        group size=3 by 1,
        xticklabels at=edge bottom,
        horizontal sep=0pt
    },
    ybar stacked, ymin=0,
    xticklabel style={/pgf/number format/1000 sep=,rotate=60,anchor=east,font=\scriptsize},
    width=\columnwidth,
    height=5cm,
    grid=major,
    xtick=data,
    ytick={5, 10},
]
\pgfplotsset{minor grid style={dashed,gray}}
\pgfplotsset{major grid style={dotted,gray!50!black}}

\pgfplotsset{legend style={at={(0.32,-0.5)},anchor=south, legend columns=-1,/tikz/every even column/.append style={column sep=0.2cm}, font=\footnotesize}
}
\pgfplotstableread[row sep=\\,col sep=&]{
year & n & c & j & w & s\\
2005 & 2 & 0 & 2 & 0 & 0\\
}\one

\pgfplotstableread[row sep=\\,col sep=&]{
year & n & c & j & w & s\\
2010 & 1 & 0 & 1 & 0 & 0\\
}\two

\pgfplotstableread[row sep=\\,col sep=&]{
year & n & c & j & w & s\\
2015 & 1 & 1 & 0 & 0 & 0\\
2016 & 0 & 0 & 0 & 0 & 0\\
2017 & 1 & 0 & 0 & 1 & 0\\
2018 & 0 & 0 & 0 & 0 & 0\\
2019 & 4 & 0 & 4 & 0 & 0\\
2020 & 13 &	2 &	8 &	0 &	3\\
2021 & 10 & 2 & 7 &	0 &	1\\
2022 & 10 & 2 & 6 &	0 &	2\\
2023 & 1 & 1 & 0 & 0 & 0\\
}\three

\nextgroupplot[xmin=2004,xmax=2006,
               xtick={2005},
               width=2.8cm,
               ymax=15,
               axis x discontinuity=parallel,
               ytick pos=left,
               yticklabel pos=left,
               axis y line = left,
               axis line style={-},
               ]
  \addplot [fill=blue!70!black] table[x=year,y=c] {\one};
  \addplot [fill=red!70!black] table[x=year,y=j] {\one};
  \addplot [fill=green!70!black] table[x=year, y=w] {\one};
  \addplot [fill=yellow!40!white] table[x=year,y=s] {\one};

\nextgroupplot[xmin=2009,xmax=2011,
               xtick={2010},
               width=2.8cm,
               ymax=15,
               axis x discontinuity=parallel,
               hide y axis,
               grid=major,
               ]
 
 \draw[dotted,gray!50!black](axis cs:2009,5) -- (axis cs:2014,5); 
 \draw[dotted,gray!50!black](axis cs:2009,10) -- (axis cs:2014,10); 
 
  \addplot [fill=blue!70!black] table[x=year,y=c] {\two};
  \addplot [fill=red!70!black] table[x=year,y=j] {\two};
  \addplot [fill=green!70!black] table[x=year, y=w] {\two};
  \addplot [fill=yellow!40!white] table[x=year,y=s] {\two};

 \nextgroupplot[xmin=2014,xmax=2024,
                ymax=15,
                width=7cm,
                axis x discontinuity=parallel,
                ytick pos=right,
                yticklabel pos=right,
                axis y line = right,
                axis line style={-},
                ]
\addplot [fill=pt_blue] table[x=year,y=c] {\three};
\addplot [fill=pt_cyan] table[x=year,y=j] {\three};
\addplot [fill=pt_teal] table[x=year, y=w] {\three};
\addplot [fill=pt_orange] table[x=year,y=s] {\three};
\legend{Journals\hspace*{1em},Conferences\hspace*{1em},Workshops\hspace*{1em}, Symposium}
\end{groupplot}
\end{tikzpicture}
\caption{Publication venues trend over the years}
\label{fig:trend}
\end{figure}
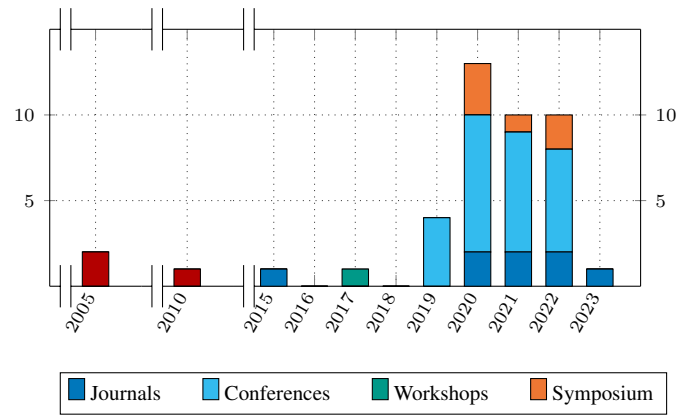
\begin{table}
    \centering
    \caption{Top 10 papers with the most citations}
    \label{tab:citations}
    \begin{tabularx}{\columnwidth}{@{}cXcc@{}}
    \toprule
    \textbf{Paper}  &  \textbf{Title} & \textbf{Year} & \textbf{Citations} \\
    \midrule
\cite{aeberhard:15} & Experience, Results and Lessons Learned from \textbf{Automated Driving} on Germany's Highways & 2015 & 302 \\
\cite{jha:19} & Networked embedded acoustic processing system for smart building applications & 2019 & 109  \\
\cite{biondi:19} & A Safe, Secure, and Predictable Software Architecture for Deep Learning in Safety-Critical Systems & 2020 & 37  \\
\cite{velasco:20} & \textbf{Autonomous Driving} Architectures, Perception and Data Fusion: A Review & 2020 & 36 \\
\cite{lotz:19} & Microservice Architectures for \textbf{Advanced Driver Assistance Systems}: A Case-Study & 2019 & 33 \\
\cite{bucaioni:20} & Technical Architectures for \textbf{Automotive Systems} & 2020 & 26 \\
\cite{serban:19} & Designing Safety Critical Software Systems to Manage Inherent Uncertainty & 2019 & 25 \\
\cite{hernandez:20} & SELENE: Self-Monitored Dependable Platform for High-Performance Safety-Critical Systems & 2020 & 22  \\
\cite{roy:21} & Micro-Safe: Microservices- and Deep Learning-Based Safety-as-a-Service Architecture for 6G-Enabled \textbf{Intelligent Transportation System} & 2022 & 21\\
\cite{biondi:21} & SPHERE: A Multi-SoC Architecture for Next-Generation Cyber-Physical Systems Based on Heterogeneous Platforms & 2021 & 15\\
    \bottomrule
    \end{tabularx}
\end{table}

\section{Discussion}
\label{sec:discussion}
Our research has revealed that primary studies focusing on AI in mobility systems are relatively scarce compared to those with other foci. To address the research questions, we analysed 38 studies, from which we derived several key findings:

\textbf{Combatting Uncertainty with Redundancies and Services.} 
Our analysis indicates that redundancy is commonly employed in both AI components and other system parts to enhance safety in case of failures or errors. Additionally, more than one-third of the papers used layered or service-oriented architectures, which likely enhances the reliability and safety of AI systems by segregating AI components from the rest of the system.

\textbf{Prevalence of Informal Notation.} 
Despite the availability of established modelling languages like UML and SysML, more than half of the studies opted for informal architectural representations. We advocate for adopting standardised modelling languages to better articulate architectures and facilitate clearer communication between researchers and implementers.

\textbf{Underutilisation of AI in Spacecraft and Railway Sectors.} 
Our study identified a minimal application of AI within the spacecraft and railway sectors. These domains represent underexplored areas where AI could potentially drive significant optimisations and improvements, particularly in railways.

\textbf{Narrow Focus of Software Architecture Solutions.}
Most architectural solutions are tailored exclusively to mobility systems with machine learning components, revealing a research gap in areas like reinforcement learning or end-to-end learning.

\textbf{Limited Validation and Evaluation of Proposed Solutions.}
Except for a few instances involving DNN and some AI and ML applications, most proposed solutions lack thorough validation. While simulation is a common validation method, there is a notable deficiency in evaluations conducted on real, industry-specific systems.

\textbf{Increasing Integration of AI in Mobility Systems.}
There is a discernible trend towards integrating AI into mobility systems, reflecting its growing significance and potential permanence within these systems.
\subsection{Threats to Validity}
Despite our systematic approach, subjectivity remains a challenge. To mitigate potential bias, the study involved both a research associate and a senior researcher, ensuring each paper was reviewed at least twice. Some relevant studies might have been overlooked due to limitations in our search terms or the databases used, although we employed well-known databases and crafted our search terms to encompass key keywords, abbreviations, and synonyms. We acknowledge that new studies might have been published during our data collection; hence, we have documented when and how we collected our data to allow for future replication of our findings.
\section{Conclusion}
\label{sec:conclusion}
In this paper, a state-of-the-art overview of existing architectural solutions in AI-based mobility systems is presented. We conducted a systematic mapping study in which we searched 1,520 papers from different databases and selected the relevant 38 papers through a voting process based on the exclusion criteria. We extracted the key data from these papers to answer the research questions. These answers provided a clear understanding of the area. 

We concluded that despite the growing trend of AI, the need for new, innovative and, above all, safe architectural solutions is far from being met. It has been shown that existing solutions have not been sufficiently validated, and therefore, the proposed approaches undermine their use in industry. We, therefore, call for an industry-oriented evaluation of the approaches and their implementation in real systems. Regarding AI, it is evident that there is still hesitation towards utilising atypical techniques like end-to-end or reinforcement learning, hindering the possible advantages of these methods. As with any new approach, there are also challenges, which is why we have tried to provide an overview of the safety patterns and frameworks in use so that future researchers are aware of the current state of the art. We also suggest not only concentrating on the automotive sector but also venturing beyond the borders and considering railway or aircraft vehicles. 

For future work, we want to expand our research area on the one hand and, for instance, analyse the same aspect in different areas such as medicine. On the other hand, we plan to analyse individual areas, such as vehicles and the architectures used in them, in much more depth.

\bibliographystyle{IEEEtranS}
\balance
\bibliography{bibliography}

\end{document}